\documentclass[preprint2,times,tighten]{aastex61}

\usepackage{epsfig}			
\usepackage{graphicx,color}		
\usepackage{amssymb}			
\usepackage{color}			
\usepackage{url}			
\usepackage{amsmath}			
\usepackage{rotating}			
\usepackage{float}			
\usepackage{textcomp}			
\usepackage{epstopdf}
\usepackage{dcolumn}
\usepackage{times}
\usepackage{tabularx}
\usepackage[english]{babel}

\newcommand{\Fig}[1]{Figure~\ref{#1}}
\newcommand{\Sec}[1]{Section~\ref{#1}}

\shorttitle{Latitude distribution of Sunspots}
\shortauthors{Sudip Mandal et al.}

\begin{document}


\title{Latitude Distribution of Sunspots: Analysis Using Sunspot Data and A Dynamo Model}

\correspondingauthor{Sudip Mandal}
\email{sudip@iiap.res.in}

\author[0000-0002-7762-5629]{Sudip Mandal}
\affil{Indian Institute of Astrophysics, Koramangala, Bangalore 560034, India}

\author{Bidya Binay Karak}
\affil{High Altitude Observatory, National Center for Atmospheric Research, 3080 Center Green Dr., Boulder, CO 80301, USA}

\author{Dipankar Banerjee}
\affil{Indian Institute of Astrophysics, Koramangala, Bangalore 560034, India}
\affil{Center of Excellence in Space Sciences India, IISER Kolkata, Mohanpur 741246, West Bengal, India}


\begin{abstract}
In this paper, we explore the evolution of sunspot latitude distribution and explore its relations with the cycle strength.  
With the progress of the solar cycle the distributions in two hemispheres from mid-latitudes propagate toward the equator
and then (before the usual solar minimum) these two distributions touch each other. 
By visualizing the evolution of the distributions 
in two hemispheres, we separate the solar cycles by excluding this hemispheric overlap. From these
isolated solar cycles in two hemispheres, we generate latitude distributions for each cycle, starting from cycle 8 to cycle 23.
We find that the parameters of these distributions, namely, the central latitude ($C$), width ($\delta$) and height ($H$)
evolve with the cycle number and they show some hemispheric asymmetries.
Although the asymmetries in these parameters persist for a few successive cycles, they get corrected within a few cycles and the new asymmetries appear again.
In agreement with the previous study, we find that distribution parameters are correlated with the strengths of the cycles,
although these correlations are significantly different in two hemispheres. The general trend
that stronger cycles begin sunspot eruptions at relatively higher latitudes and have wider bands of sunspot emergence latitudes are confirmed
when combining the data from two hemispheres.
We explore these features using a flux transport dynamo model with stochastic fluctuations.
We find that these features are correctly reproduced in this model.
The solar cycle evolution of the distribution center is also in good agreement with observations.
Possible explanations of the observed features based on this dynamo model are presented.
\end{abstract}

\section{Introduction}
A beautiful feature of the solar cycle is its latitude-time distribution,
so-called the butterfly diagram.
At the beginning of the cycle, 
sunspots appear to be distributed around the mid-latitudes and with the progress
of the cycle, the distribution moves towards the equator. 
This is known as the equatorward migration of sunspots \citep{2011SoPh..273..221H}.
This feature of solar cycle has become a central interest to many solar astronomers as well as to dynamo modelers.
As the strengths are not the same for all the cycles, sunspot latitudinal distributions are also expected to be different.
However, it is found that the cycle strength has some relations with the properties of these distributions.
\citet{2000JApA...21..163S} identified a correlation between the cycle strength and the mean latitude of sunspot distribution.
They showed even a stronger correlation between the width and the mean latitude of such distribution.
Later, using the group sunspot data, \citet{2003SoPh..215...99L} showed that
the number of sunspot group present at latitudes $\geq$ 35\textdegree\
is positively correlated with the amplitude of the cycle \citep[also see][]{2017A&A...599A.131L}.
Finally, \citet{2008A&A...483..623S} have computed various moments of sunspot distribution, separately for two hemispheres,
and have shown that the three lowest moments (i.e, the total area covered by the sunspots over a cycle, the mean sunspot latitudes, and the width of the distribution) are well correlated with each other.
Recently, \cite{IM16} showed that the latitude properties of sunspot distribution are much
more stable against the gaps of observations and hence these properties can be used for estimates of
quality of observations and for data series calibration.

Based on the polarities of the bipolar sunspot \citep[Hale polarity rule;][]{Hale19}, it is believed that the
sunspots are produced from the toroidal (east-west directed) field underneath the surface. The sign of this toroidal field
must be opposite in two hemispheres during a cycle and the polarity flips in every cycle.
Thus the propagation of the distribution of the sunspots at the surface is an observable of the equatorward migration of the toroidal field band in the interior of the Sun.
 For this equatorward migration, two major explanations are available in the literature. 
One is due to the dynamo wave whose direction of propagation is determined by signs of radial shear and $\alpha$ (so-called the Parker-Yoshimura sign rule)
and other is due to the equatorward drift of the toroidal flux by an equatorward flow (meridional circulation and/or magnetic pumping).
Based on theories and observations of the differential rotation, $\alpha$ effect, and the meridional circulation, we expect that the equatorward drift of the toroidal field by the flow is the cause of the propagation of the sunspot distribution in the Sun,
although we do not want to exclude the other possibility and, in fact, we do not have to debate this issue here \citep{KC16, CDB17}.

Recently, \citet{CS16} have identified the equatorward migration of sunspot-producing toroidal field band
from the equatorward migration of sunspot distribution \citep[see also][]{2011SoPh..273..221H, Jiang11}.
In each hemisphere, they have approximated sunspot latitudes for every year with a Gaussian distribution.
For all cycles, \citet{CS16}  have shown that this Gaussian in each hemisphere evolves separately, and depending on the strength of the
cycle, the activity and the width of the Gaussian increase first and then decline at the same rates.
They interpreted this result with the fact that in the solar interior two distinct toroidal flux bands in two hemispheres
are transported towards the equator and when these two bands begin to touch each other they cancel. This
cancellation and the cross-equator diffusion of toroidal magnetic flux, which begin a few years before the usual solar minimum,
are the processes in which the solar cycle interacts with the other hemisphere. In other words, there is a hemispheric overlap
at the end of each cycle in contrast to the usual inter-cycle overlap as emphasized by \citet{2007ApJ...659..801C}.

In this manuscript, we isolate the solar cycle in each hemisphere by excluding the cancellation and
the cross-equator diffusion of toroidal magnetic flux in two hemispheres by removing the hemispheric overlap
and study the properties of the latitudinal distribution of the solar cycle.
In previous studies \citep{2000JApA...21..163S, 2008A&A...483..623S, 2003SoPh..215...99L}, the solar cycles were separated only based on the usual inter-cycle
overlap, and not the hemispheric overlap. We shall explore how the properties of sunspot distribution
of the isolated solar cycle evolve in two hemispheres.
Next, to study these features, we shall employ a Babcock--Leighton type flux transport dynamo
model in which the toroidal flux near the bottom of the convection zone, which is
assumed to produce sunspots at the surface, is transported towards the equator
by an equatorward meridional flow. This dynamo model
has become a popular paradigm in recent years for modeling the solar cycle at present \citep{Kar14a,Ch15}.
We show that by including fluctuations in the Babcock--Leighton process of this model, all the observed properties can be correctly reproduced.

\section{ Observational Data }\label{sec:data}

In our study, we mainly use the RGO sunspot data\footnote{\url{https://science.nasa.gov/ssl/pad/solar/greenwch.htm}}.
This data, however, covers only from cycle 12 to cycle 23 (year 1874--2011) and thus
to obtain the data for previous cycles, i.e., cycles 7--11, we use the recently digitized sunspot catalog 
compiled by \citet{2017A&A...599A.131L}; see CDS\footnote{\url{http://vizier.cfa.harvard.edu/viz-bin/VizieR?-source=J/A+A/599/A131}}. 
This data only provides locations of sunspots from the digitized drawings of Schwabe (1826--1867) and Sp{\"o}rer (1866--1880).
Since the early years of cycle 7 is not available in the \citet{2017A&A...599A.131L} catalog,
we shall begin our analysis from cycle 8.
Although this new catalog contains data for up to present cycle, for cycles 12--23 we still use RGO data which is popular and has been used in many previous studies.
We also make use of the sunspot number data provided by the Solar Influences Data Center (SIDC). 
This is the newly calibrate data\footnote{\url{http://www.sidc.be/silso/datafiles}} (V2.0). 

\begin{figure}
\centering
\includegraphics[width=0.5\textwidth]{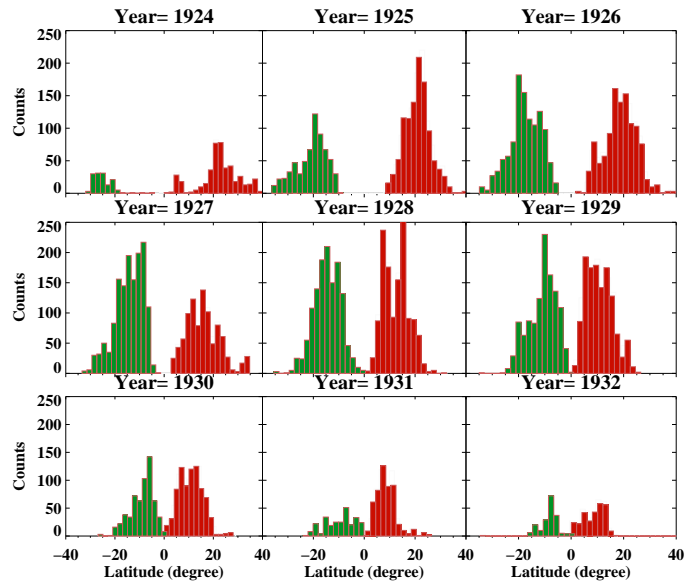}
\caption{Latitude distributions of sunspots from 1924--1932 i.e., covering most of the solar cycle 16. A bin size of $2^\circ$ is used to compute the distributions. Red and green colors represent the northern and southern hemispheres, respectively.}
\label{overlap_context}
\end{figure}



\section{Methods}\label{sec:method}
As discussed in the Introduction,
 our interest is to investigate the properties of sunspot latitude distribution 
of individual hemispheric cycle. Thus, for a given cycle, we restrict our analysis between the start of the cycle and 
the year when the two latitude distributions of two hemispheres come in contact with each other. 
This definition is motivated by \citet{CS16} and it allows us to capture the behavior 
of the isolated cycle, without having any influence from the other hemisphere. 
In our analysis, first we remove the inter-cycle overlap \citep{2007ApJ...659..801C}, by excluding one year
at the beginning and one year at the end of the cycle.
Next for every one-year data, we make the distribution of the sunspot latitudes
as shown in \Fig{overlap_context} for cycle~16.
We observe that the separation between the two distributions is large 
in the beginning of the cycle (year 1924) and progressively this becomes narrower.
Eventually two distributions come in contact with each other (in year 1929). 
Thus, the duration of the cycle 16 is considered as the interval between 1924--1929. 
The above procedure is repeated for all the other cycles. 
Identification of the year, when the two distributions come in contact 
with each other, has been done visually. 

\begin{figure}
\centering
\includegraphics[width=0.5\textwidth]{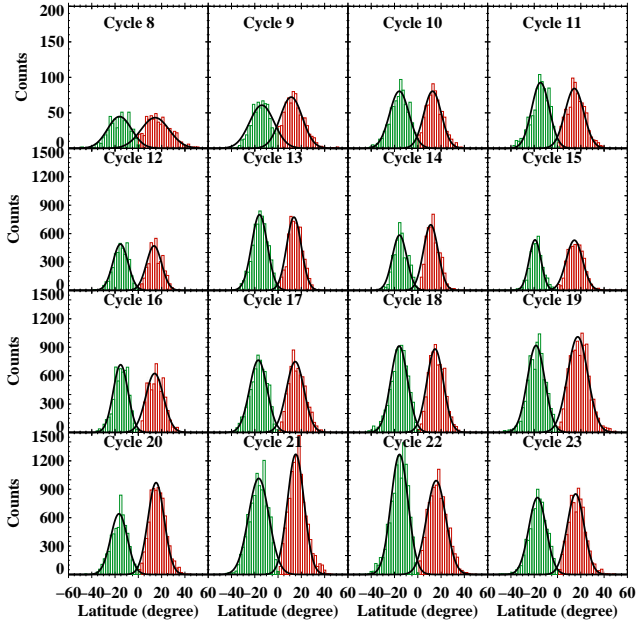}
\caption{Histograms showing the latitude distributions of the sunspots for cycles 8--23. Red and Green colors highlight the northern and southern hemispheres. Black solid curves represent the fitted Gaussian functions on the individual hemispheric distributions.} 
\label{gauss_fit}
\end{figure}

Next, for each hemisphere we generate the combined latitude distributions 
by taking the data within the cycle duration as identified as above.
These distributions for cycles 8--23 are displayed in \Fig{gauss_fit}. 
We fit the hemispheric distributions with two individual Gaussians as shown with the solid black lines in every panel in \Fig{gauss_fit}.
From every fitted Gaussians, we note three parameters: Height $H_{\rm{ gauss}}$ (in arbitrary unit), 
Central latitude $C_{\rm{ gauss}}$ (in degrees), and the width $\delta_{\rm{ gauss}}$ (in degrees).  
In addition, we also calculate the mean $C_{\rm{ moment}}$, standard deviation $\delta_{\rm{ moment}}$,
and the height of the distribution at the distribution mean $H_{\rm{ moment}}$
directly from original distributions. 
We note that when the distribution is perfectly Gaussian, 
parameters ($C$, $\delta$, and $H$) obtained from these two methods must be identical.

\begin{figure*}[!htb]
\centering
\includegraphics[width=0.95\textwidth]{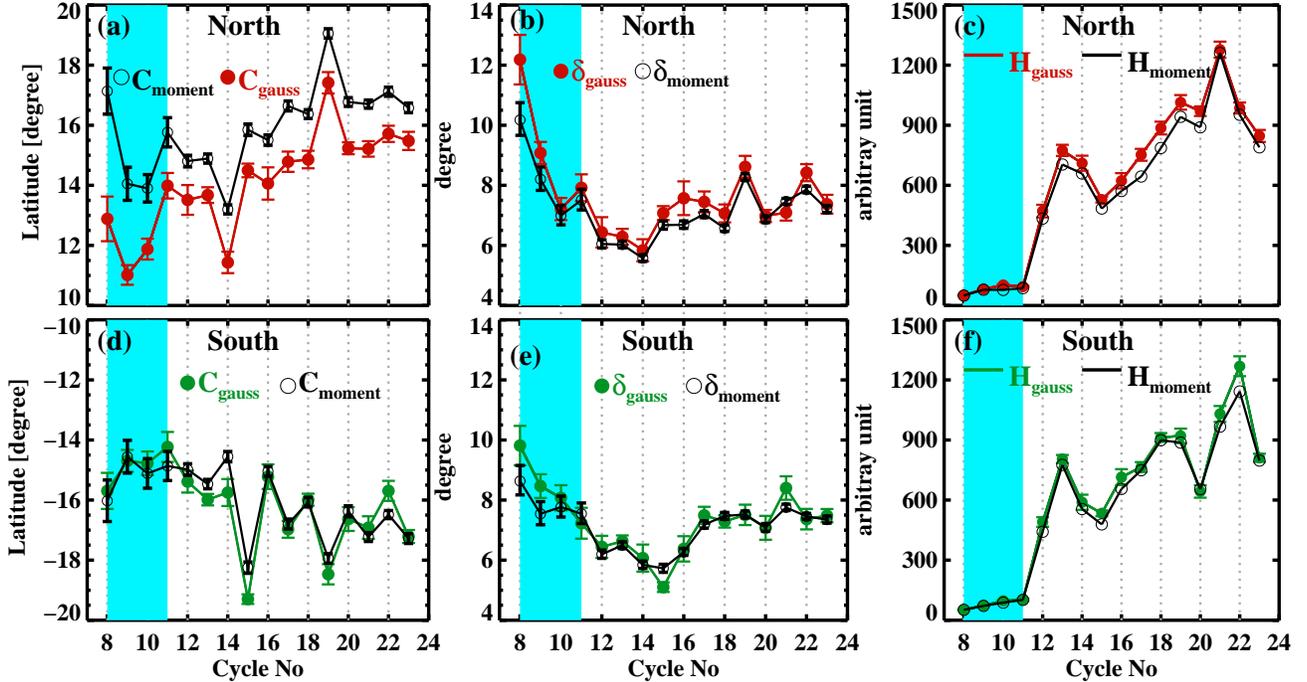}
\caption{Left to right: Evolutions of $C$, $\delta$, and $H$ computed from the latitude distributions (open circles) 
and from the fitted Gaussians (filled circles). Top and bottom panels are obtained from northern and southern hemispheres, respectively.
The shaded region shows the data obtained from \citet{2017A&A...599A.131L}, while the unshaded region shows the result from RGO.
}
\label{param_evolve}
\end{figure*}

\section{Results}
\subsection{Evolutions of distribution parameters with solar cycle}\label{sec:evol}
In \Fig{param_evolve}, we display the evolutions of hemispheric $C$, $\delta$, and $H$ parameters 
as obtained from the fitted Gaussian (filled points). The same parameters obtained from the distribution moments are 
also displayed (open circles) for comparison. 
The error bars on the Gaussian parameters represent the one sigma errors whereas, for the distribution moments, 
it represents the 95\% confidence interval of the derived quantity. 
We notice that the solar cycle variations of these parameters obtained from two methods
although follow similar patterns, there are some differences in many cycles. 
Interestingly, we see much deviations in the northern hemisphere as compared to the southern one. 
Seeing these significant differences we can guess that the Gaussian profile does not fit the sunspot distribution adequately.
To check the significance of the fitted Gaussian, we 
perform a normality test,`Shapiro-Wilk test' to all observed latitude distributions and
the results are presented in Table~\ref{normality_test}. 
This test as proposed by \citet{doi:10.1093/biomet/52.3-4.591}, 
calculates the $\rm{W}$ statistic whether a random sample comes from a normal distribution. 
Higher $\rm{P}$ value ($\geq0.05$) along with a high $\rm{W}$ ($>0.99$) indicate that the `null-hypothesis' 
(that the sample is drawn from a normal distribution) cannot be rejected. 
We note that test is also applicable to a small sample size \citep{Royston1992}. 
From the tabulated values, we note that only a few cycles (cycles 10, 11, 12, and 14) 
are having $\rm{P}$ $\geq0.05$. Values of $\rm{W}$ are always lower than the critical value for all cycles.
We thus conclude that the latitude distributions cannot be faithfully described by a Gaussian profile.

\begin{figure*}
\centering
\includegraphics[width=0.98\textwidth]{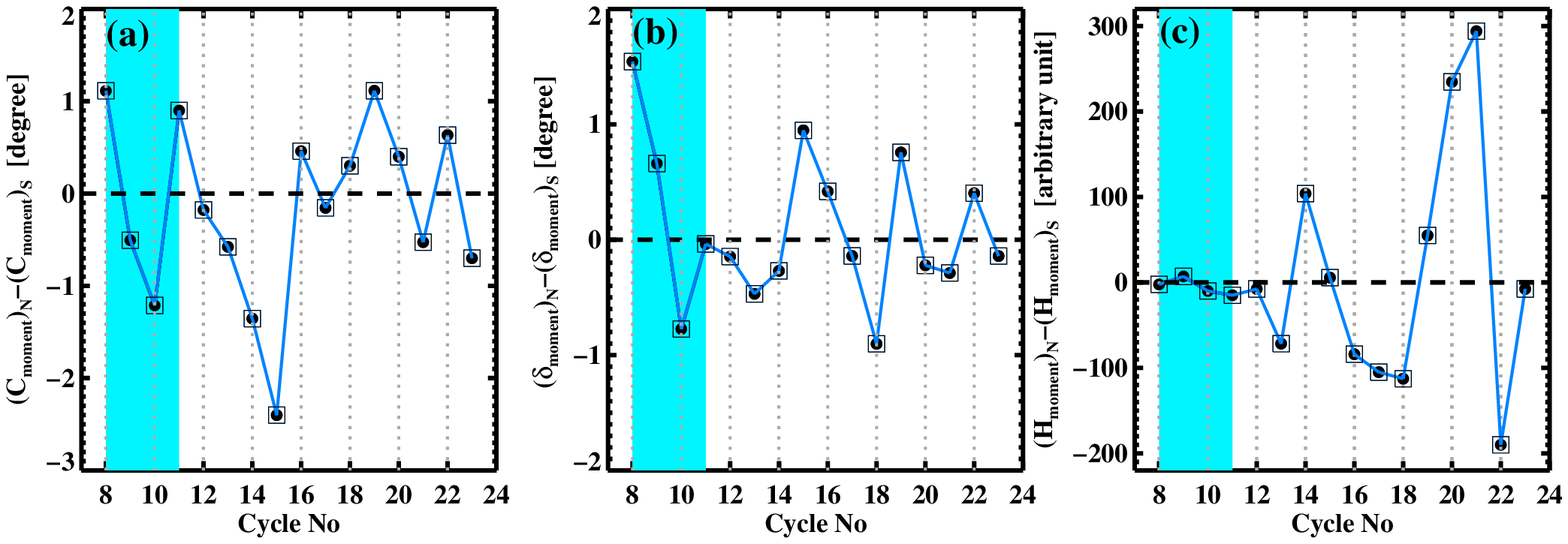}
\caption{Different panels show the north-south asymmetry in the $C_{\rm{ moment}}$, $\delta_{\rm{ moment}}$, and $H_{\rm{ moment}}$ parameters.} 
\label{n_s_asym}
\end{figure*}

\begin{figure}
\centering
\includegraphics[width=0.35\textwidth]{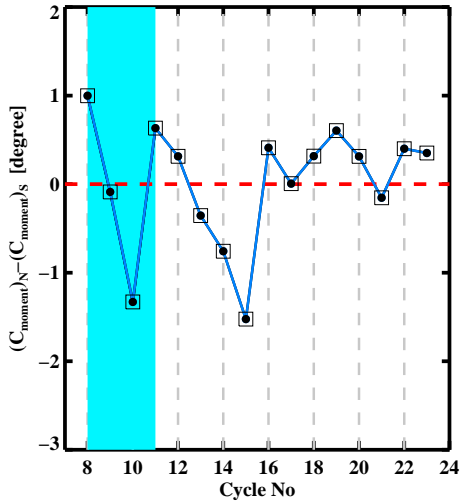}
\caption{Same as \Fig{n_s_asym}a, but in this case, 
the cycle length is computed from the interval between cycle beginning and end,
i.e., it does not exclude the hemispheric overlap.}
\label{n_s_asym_yearly}
\end{figure}

We would like to mention that despite poor Gaussian fitting of the sunspot latitude distribution,
many previous authors have used Gaussian distribution.
Particularly, \citet{2017A&A...601A.106M} have shown that the latitude distribution of sunspots, in each hemisphere, 
integrated over the cycle length, follows a Gaussian distribution. 
\cite{CS16}, in fact, have used a Gaussian profile for the yearly sunspot data.
There are also attempts to use a double-Gaussian model to express the latitude distributions\citep{2012NewA...17..247C}. 
Whereas \citet{2013Ge&Ae..53..962M} have shown a significant skewness in the distribution 
and this skewness varies over the solar cycle.
However, as in our study, we can obtain the required parameters of the sunspot distributions directly from the latitude distributions,
we shall only use $C_{\rm{ moment}}$, $\delta_{\rm{ moment}}$, and $H_{\rm{ moment}}$ for our further analysis. 
Thus our computation of these parameters is essentially similar to that of \citet{2008A&A...483..623S}.

Returning to the solar cycle variations of sunspot latitude distribution 
in \Fig{param_evolve}, we find that $C$ and $H$ follow increasing trends with the cycle number. 
The most interesting feature we observe is the hemispheric asymmetry. 
For first few cycles, the variation of C is very different in two hemispheres.
In the southern hemisphere, C does not increase much with the cycle number. 
Also, for $C_S$ there is a large peak in cycle 15, while for $C_N$ the largest peak is in cycle 19. 
Therefore when we compute the difference of $C_{\rm{moment}}$ between two hemispheres, 
we obtain a nonzero value for most of the cycles; see \Fig{n_s_asym}(a). 
The interesting behavior we see is that the asymmetry does not follow any particular pattern. 
For some cycles C remains closer to the equator in the northern hemisphere and then suddenly 
C becomes closer to the equator in the southern hemisphere.
Although for most of the cycles the asymmetry persist for more than one cycle, it gets corrected randomly and thus no hemisphere dominates permanently.

We mention that our result is very similar to the one obtained in \citet{1999A&A...341L..43P} 
and \citet{2010AN....331..765Z} who presented the asymmetry of $C$, although the comparison is 
not straightforward. We recall that in the previous analyses, cycle lengths are computed from 
cycle minimum to minimum, while in our analysis the length is defined by excluding the hemispheric 
overlap of sunspot distributions. Therefore, to check how our $C_N-C_S$ gets affected with 
the definition of the cycle length, we repeat our analysis by defining the cycle length as 
done in \citet{1999A&A...341L..43P} and the result is shown in \Fig{n_s_asym_yearly}.
In this figure, we see a different variation. For a few cycles, the sign of $C_N-C_S$ has 
changed as compared to the one found in \Fig{n_s_asym}(a), although the overall variation is not 
very different. Keeping in mind that our analysis contains a longer data and $C_N$ and $C_S$ are 
computed over one-cycle data, the variation of C$_N$$-$C$_S$ obtained in \Fig{n_s_asym_yearly} is 
equivalent to that found in Fig.\ 4 of \citet{1999A&A...341L..43P}.


 \begin{table}
\begin{center}
\centering
\begin{scriptsize}
\caption{Details from the Shapiro-Wilk test}  
\vspace{0.5cm}
\label{normality_test}
\begin{tabular}{lccccccr} 
  \hline
   \multicolumn{1}{c}{Cycle}& \multicolumn{1}{c}{Hemi-} & \multicolumn{1}{c}{$\rm{P}$} & \multicolumn{1}{c}{$\rm{W}$} & \multicolumn{1}{c}{Hemi-} & \multicolumn{1}{c}{$\rm{P}$} & \multicolumn{1}{r}{$\rm{W}$} \\
   \multicolumn{1}{c}{}& \multicolumn{1}{c}{sphere} & \multicolumn{1}{c}{$\rm{}$} & \multicolumn{1}{c}{$\rm{}$} & \multicolumn{1}{c}{sphere} & \multicolumn{1}{c}{$\rm{}$} & \multicolumn{1}{r}{$\rm{}$} \\
     
     \hline
       8 & N & 0.0397 & 0.9169 & S & 0.0255 & 0.9066\\
       9 & N & 0.0024 & 0.8653 & S & 0.0481 & 0.9084\\
      10 & N & 0.1027 & 0.9196 & S & 0.0678 & 0.9156\\
      11 & N & 0.1006 & 0.9191 & S & 0.0360 & 0.8964\\
      12 & N & 0.1079 & 0.9150 & S & 0.0062 & 0.8571\\
      13 & N & 0.0027 & 0.8456 & S & 0.0130 & 0.8806\\
      14 & N & 0.0023 & 0.8422 & S & 0.0544 & 0.8985\\
      15 & N & 0.0003 & 0.7357 & S & 0.0001 & 0.7086\\
      16 & N & 0.0238 & 0.8872 & S & 0.0065 & 0.8467\\
      17 & N & 0.0109 & 0.8844 & S & 0.0240 & 0.8972\\
      18 & N & 0.0002 & 0.8173 & S & 0.0075 & 0.8770\\
      19 & N & 0.0010 & 0.8485 & S & 0.0063 & 0.8770\\
      20 & N & 0.0010 & 0.8344 & S & 0.0073 & 0.8800\\
      21 & N & 0.0181 & 0.8946 & S & 0.0009 & 0.8544\\
      22 & N & 0.0090 & 0.8870 & S & 0.0010 & 0.8484\\
      23 & N & 0.0030 & 0.8625 & S & 0.0034 & 0.8683\\

  \hline

\end{tabular}
\end{scriptsize}
\end{center}
\end{table}

Now we go back to \Fig{param_evolve}(b) to follow the evolution of $\delta$ in the two hemispheres. 
We find that for cycle 8, $\delta$ is significantly higher compared to all other cycles.
One possible reason for this could be the sparse data during this cycle, resulting larger error bar too. 
We also find a significant hemispheric asymmetry in $\delta$ as shown in \Fig{n_s_asym}(a). 
In this case, $\delta _{N}-\delta _{S}$ shows some oscillations around zero.
However, given the limited data, we cannot confirm any periodicities that might exist in $\delta _{N}-\delta _{S}$. 
Interesting, this asymmetry tends to decrease with the cycle number.

In \Fig{param_evolve}(c), the variation of $H$ parameter shows a similar trend in two hemispheres except for a few cycles. 
Particularly, $H$ has the highest peak for cycle~21 in the northern hemisphere, while in the southern hemisphere the highest peak is in cycle 22.
Similar to other parameters, we do find a considerable hemispheric asymmetry. However, unlike $\delta$, $H$ shows higher asymmetry
during recent cycles.

\subsection{Correlations between cycle strength and distribution parameters}
\subsubsection{Considering hemispheric data}
\begin{figure*}
\centering
\includegraphics[width=0.95\textwidth]{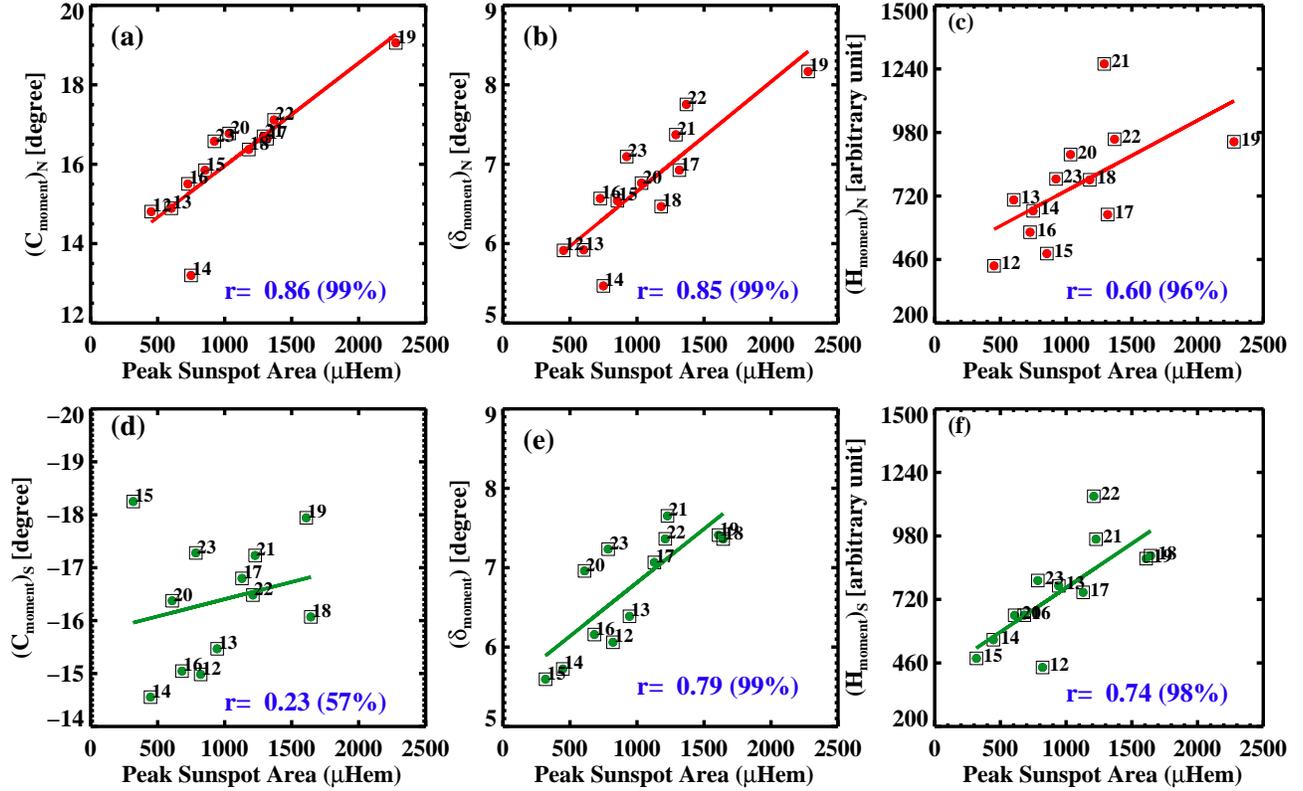}
\caption{ Left to right: scatter plots of $C$, $\delta$, and $H$ with the peak sunspot area. 
The top and bottom (green) panels represent the values obtained for the northern and southern 
hemispheres respectively. The linear Pearson correlation coefficients (r) between parameters 
(along with the confidence levels) are printed on each panel.} 
\label{obs_ns_corr}
\end{figure*}

We now demonstrate the dependence of these distribution parameters with the cycle strength. 
In \Fig{obs_ns_corr}, we show the scatter plots of $C_{moment}$, $\delta_{moment}$, and $H_{moment}$ with the peak sunspot area.
The linear Pearson correlation coefficients are also displayed in these plots. 
Correlations using the Gaussian parameters are also computed and are 
summarized in Table~\ref{cc_comp} for the comparison.
We note that in these correlations plots, we have used the peaks of yearly sunspot area data as measures of the cycle strengths 
because no homogeneous, long-term hemispheric sunspot number data is available before 1992. 
We also note that in the correlation plots, we do not consider the data of  \citet{2017A&A...599A.131L} for cycles 8--11.
We already noticed in \Fig{param_evolve} that although the \citet{2017A&A...599A.131L} catalog 
provides data from cycle 8, this data is very different than RGO data. We realize that 
the number of spots for a given day is different than that provided by the RGO record.
Although $C$ and $\delta$ parameters are less affected by such discrepancies, the 
$H$ parameter, which has a strong dependency on the number of spots, is significantly affected 
(see the sudden increase in \Fig{param_evolve}(c) for $H$ in cycle~12). Therefore, if we include cycles 8--11 data 
in our correlation analysis, then some of the results get spoiled; see Table~\ref{cc_comp}. 
Furthermore, since we do not have systematic area measurements prior to cycle 12, computation of correlations in this case are restricted to cycles 12--23. 

In \Fig{obs_ns_corr} we observe a significant positive correlation (r = 0.86) 
between $C_N$ and the peak sunspot area. However, the same correlation is insignificant 
in the southern hemisphere (r = 0.23). This correlation does not improve when computed using the $C_{\rm{ gauss}}$ data (Table~\ref{cc_comp}).
Even when this correlation is computed using the data of \citet{2017A&A...599A.131L} for all the cycle, the correlation is weaker in the southern hemisphere.

We note that our result is in agreement with \cite{2017A&A...599A.131L} who also find a weak correlation 
between the first sunspot latitudes (they call it the first latitude) and the wing strengths and no correlation between the latitudes of the last sunspot groups (end of the wing) and the wing strengths.
Interestingly, \citet{2008A&A...483..623S} found a stronger correlation between the total sunspot area and $C_{\rm{moment}}$ 
in the southern hemisphere. Thus the correlation between $C$ and cycle strength is very sensitive on how we separate the individual cycle.
Interestingly, the correlations between $\delta_{\rm{moment}}$ and the cycle strength 
is significant in both hemispheres (\Fig{obs_ns_corr}b,e). 
Our correlation values are comparable to the ones obtained in \citet{2008A&A...483..623S}.
Slight differences are again due to the definition of the cycle length used to generate the latitude distributions here.

\begin{figure*}
\centering
\includegraphics[width=0.99\textwidth]{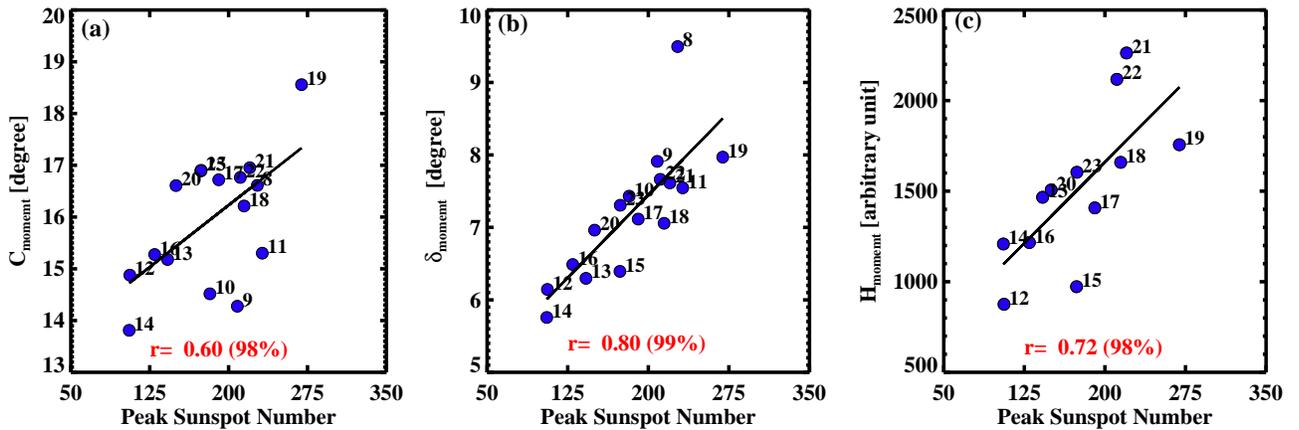}
\caption{ Same as \Fig{obs_ns_corr} but distribution parameters are obtained from the unsigned latitudes (full disc) 
and the horizontal axes represent the peak sunspot number. } 
\label{unsigned_corr}
\end{figure*}

Next, we explore what determines the higher $H$ of the latitude distribution.
There can be two reasons for having a higher $H$ of the distribution:
(i) larger number of sunspots or (ii) small $\delta$ value. 
However, when we compute the correlation between $H_{\rm{moment}}$ and $\delta_{\rm{moment}}$
we find positive correlations. The linear correlation coefficients (and confidence levels) 
are 0.64 (97\%) and 0.84 (99\%) for northern and southern hemispheres, respectively. 
However, in both hemispheres we get a positive correlation between $H_{\rm{ moment}}$ and 
the cycle strength (\Fig{obs_ns_corr}c,f). 
As discussed earlier that the spot number is different in \citet{2017A&A...599A.131L} data, this correlation completely disappears
when we consider this data; see the second row of Table~\ref{cc_comp}. 
However, based on the correlations obtained from RGO data, we confirm that stronger cycles with higher $H$ have
bigger widths of their distributions.

 \begin{table*}
\begin{center}
\centering
\caption{The linear Pearson correlation coefficients (r) 
and the confidence levels between various parameters.}  
\vspace{0.2cm}
\label{cc_comp}
\begin{tabular}{lccr}
  \hline
   \multicolumn{1}{c}{Parameters} & \multicolumn{1}{c}{Hemisphere} & \multicolumn{1}{c}{$r_{\rm{moment}}$}  & \multicolumn{1}{c}{$r_{\rm{gauss}}$} \\
     \hline
       \textbf{$\mathrm{Data~source: RGO (12-23)}$} &  &   & \\
     \hline
       $C$ vs (Sunspot Area)$_{peak}$& N & 0.86 (99\%) &  0.80 (99\%)\\
                                          & S & 0.23 (57\%) & $-$0.02 (06\%)\\
  $\rm{\delta}$ vs (Sunspot Area)$_{peak}$& N & 0.85 (99\%) &  0.78 (99\%)\\
                                          & S & 0.79 (99\%) &  0.72 (98\%)\\
  $H$ vs (Sunspot Area)$_{peak}$ & N & 0.60 (96\%) &  0.64 (97\%)\\
                                          & S & 0.74 (98\%) &  0.70 (98\%)\\
  \hline
       \textbf{$\mathrm{Data~source: Leussu~et~al (8-11)+ RGO (12-23)}$}&  &   & \\
  \hline
  $C$ vs (Sunspot Number)$_{peak}$& combined & 0.60 (98\%) &  0.33 (81\%)\\
$\rm{\delta}$ vs (Sunspot Number)$_{peak}$& combined & 0.80 (99\%) &  0.66 (99\%)\\
$H$ vs (Sunspot Number)$_{peak}$& combined & 0.72 (98\%) & 0.74 (98\%)\\

  \hline
       \textbf{$\mathrm{Data~source: Leussu~et~al (12-23)}$}&  &   &  \\
     \hline
       $C$ vs (Sunspot Area)$_{peak}$& N & 0.78 (99\%) &  0.59 (95\%)\\
                                          & S & 0.40 (83\%) &  0.19 (48\%)\\
  $\rm{\delta}$ vs (Sunspot Area)$_{peak}$& N & 0.84 (99\%) &  0.73 (98\%)\\
                                          & S & 0.58 (95\%) &  0.67 (97\%)\\
  $H$ vs (Sunspot Area)$_{peak}$ & N & $-$0.15 (39\%)  &  $-$0.07 (18\%)\\
                                          & S & 0.26 (63\%)  &  0.30 (69\%)\\
  \hline
       \textbf{$\mathrm{Data~source: Leussu~et~al (8-23)}$}&  &   &  \\
     \hline
  $C$ vs (Sunspot Number)$_{peak}$& combined & 0.60 (98\%) &  0.36 (81\%)\\
$\rm{\delta}$ vs (Sunspot Number)$_{peak}$& combined & 0.76 (99\%) &  0.67 (99\%)\\
$H$ vs (Sunspot Number)$_{peak}$& combined & $-$0.21 (64\%) & $-$0.20 (57\%)\\
  \hline
\end{tabular}
\end{center}
\end{table*}

\subsubsection{Merging the hemispheres}
We now explore whether the previous correlations obtained
from individual hemispheric parameters survive when we combine two hemispheres.
To do so, we compute distributions from the unsigned latitudes of sunspots. 
Similar to the previous analysis, we compute various moments from these distributions
and the correlation plots between the cycle strengths and these distributions parameters are 
 shown in \Fig{unsigned_corr}. We note that now we do not need hemispheric data 
and thus we take the peak sunspot number as the strength of the solar cycle, 
instead of sunspot areas that we have considered in the previous section. 
As seen in \Fig{unsigned_corr}, significant  positive correlations exist for all the parameters.
Therefore, we confirm that stronger cycles have higher mean latitudes, 
widths, and peaks of the distributions.

\begin{figure*}
\centering
\includegraphics[width=0.90\textwidth]{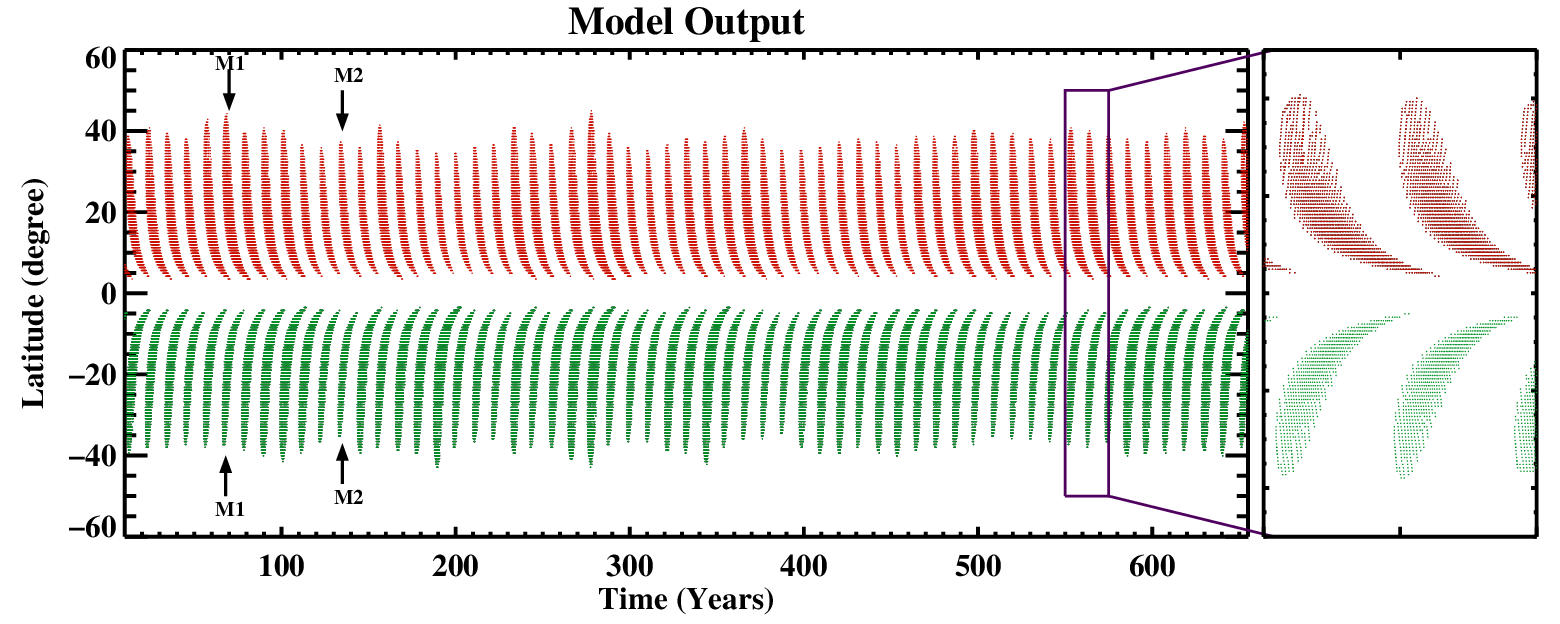}

\caption{Butterfly diagram generated using a Babcock-Leighton type flux transport dynamo model. The inset box shows a zoomed in view of the diagram for better visualization. Two arrows highlight the cycles used for generating Figure~\ref{model_histo}.} 
\label{model_butterfly}
\end{figure*}
\begin{figure}
\centering
\includegraphics[width=0.5\textwidth]{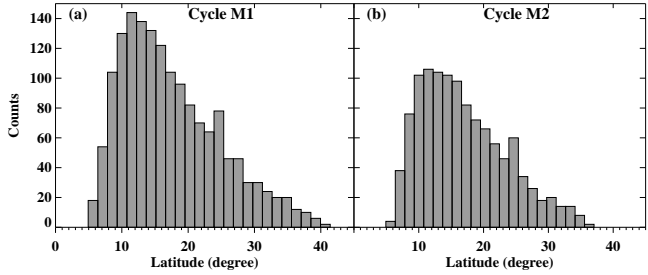}
\caption{Representative example of the latitude distributions obtained using the dynamo model (cycles are highlighted by arrows in Figure~\ref{model_butterfly}). } 
\label{model_histo}
\end{figure}
\begin{figure*}
\centering
\includegraphics[width=0.90\textwidth]{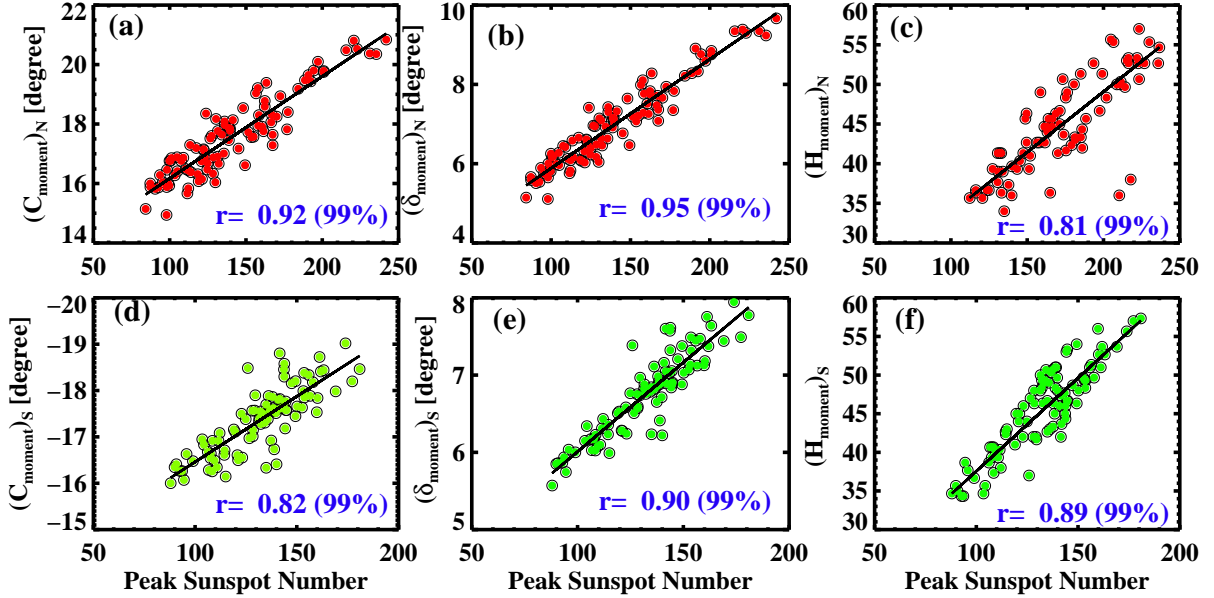}
\caption{Scatter plots showing the correlations of $C_{\rm{ moment}}$, $\delta_{\rm{ moment}}$, and $H_{\rm{ moment}}$ 
with the peak spot numbers as obtained from the dynamo model. Top and bottom panels 
are obtained from the northern and southern hemispheres, respectively.} 
\label{model_corr}
\end{figure*}
\begin{figure}
\centering
\includegraphics[width=0.5\textwidth]{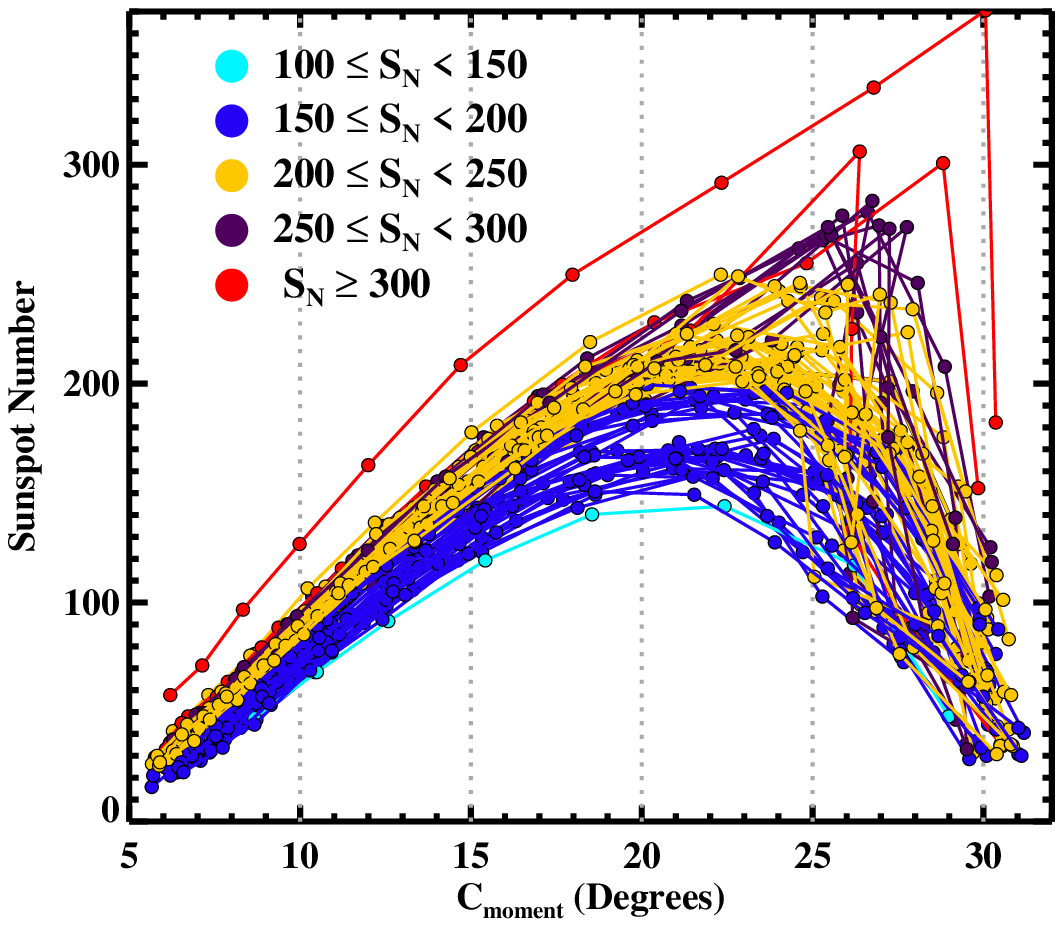}
\caption{Trajectories of the centers ($C$) of the latitude distributions of yearly spot numbers
plotted with the number of spots in each year of the cycle. 
Thus each point corresponds to one-year data and each curve corresponds to one cycle. 
The cycle begins from right and ends at the left.
Cycles are colored based on their strength (as measured by the peak spot number). These classifications are printed on the plot.} 
\label{model_cameron}
\end{figure}

\section{Results Using A Flux Transport Dynamo Model }\label{sec:dynamo}

To explore above features of the solar cycle in a theoretical model, we
consider a Babcock--Leighton type flux transport dynamo model \citep{CSD95,DC99}. 
In this study, we consider the same dynamo model as used in \citet{KC11}. The original model was developed by \citet{NC02} and \citet{CNC04} and
later this model has been used in many studies \citep[see a review by][]{Ch15}. 
Thus without providing the details of this model in the present study, 
we mention the salient features of this model.

This is an axisymmetric dynamo model in which the poloidal and toroidal components of the magnetic field are evolved with a given axisymmetric velocity field. In this model, the toroidal field is produced from the poloidal field through differential rotation in the convection zone, while the poloidal field is produced near the solar surface from the toroidal field at the base of the convection zone through an $\alpha$ effect. 
This $\alpha$ essentially captures the Babcock--Leighton process in which tilted bipolar sunspots decay and disperse to produce poloidal field on the solar surface. 
This model also includes a single-cell (in each hemisphere) meridional flow such that it is poleward in the upper half of the convection zone and equatorward in the lower half. This model reproduces the regular solar cycle when the flow and the $\alpha$ are fixed.

In this flux transport dynamo model, the poloidal flux can be changed due to multiple reasons. The main reason for this is the fluctuations in the Babcock--Leighton process through the observed variations in the tilt angle around Joy's law  \citep{Das10, SK12, JCS14, KM17}. Thus to capture the variation in the poloidal flux and to produce irregular solar cycles in this model, we follow the same procedure as prescribed by \cite{CCJ07}. At the end of every solar cycle when polar field reaches to its maximum value in this model, we change the poloidal field above $0.8R_\odot$ by a factor $\gamma$, where $\gamma$ is obtained randomly from a uniform distribution within $[0.5, 1.5]$. We do this change independently in two hemispheres because the randomness in the Babcock--Leighton process is uncorrelated in hemisphere.

We run the model for 93 cycles by changing the polar field at each solar minimum
and the butterfly diagram of {\it model sunspot} eruptions is shown in \Fig{model_butterfly}. 
As we can see from this diagram that most of the basic features of the solar cycle are reproduced in this model.
We also notice that for some cycles sunspots start appearing at somewhat higher latitudes. 
It turns out that for stronger cycles (e.g., Cycle~M1 as shown by the first arrow in \Fig{model_butterfly}) 
the first latitudes of sunspot appearance happen at relatively higher latitudes, and vice versa. 
This is in good agreement with observation as illustrated by \citet{2017A&A...599A.131L} in a detailed analysis.
However, there are some minor differences between the model and observed butterflies. 
Particularly, in the model, no spots are produced within $\pm$5$^\circ$ latitudes 
(see \Fig{model_butterfly}), while in observations, we find spots even close to the equator. 
Therefore, we have a fewer data near the equator of the histograms as shown in \Fig{model_histo} 
and thus the distributions look more skewed than the observed one.

From the model latitude distributions, we compute $C_{\rm{ moment}}$, $\delta_{\rm{ moment}}$, and $H_{\rm{ moment}}$
for all cycles in the same way as done for the observed data. 
The correlation plots between various quantities are plotted in \Fig{model_corr}. 
Strikingly, all the correlations found in observations (\Fig{obs_ns_corr} and \Fig{unsigned_corr}) are beautifully reproduced in both hemispheres.
See the correlation coefficients and confidence levels printed on the plot.
For the full disc data, the correlation coefficients (confidence levels) 
of cycle strength with $C$, $\delta$ and $H$ are 0.90 (99.9\%), 0.95 (99.9\%), and 0.88 (99.9\%), respectively.

 The correlation between the center and the peak sunspot number is understood in the following way. In our model, the poloidal field is produced near the surface at low latitudes. This field is advected towards the higher latitudes (due to the poleward flow) and then down to the bottom of the convection zone (due to the downward flow). In this process, the diffusion also contributes to transport the poloidal field from the surface to the deeper convection zone. The poloidal field then produces toroidal field through the differential rotation. This toroidal field is transported to the low latitudes by the equatorward meridional flow and produces sunspot eruptions when it exceeds a certain value. This is how we get spots below about $\pm30^\circ$ latitudes. However, due to fluctuations in the poloidal field generation process, when the toroidal field becomes strong, it starts exceeding the field strength for spot eruption at somewhat higher latitudes before reaching lower latitudes.
This allows the model to start spot eruptions at relatively higher latitudes.
Thus for the stronger cycles (with more sunspot eruptions), eruptions begin at slightly higher latitudes. This causes a positive correlation between the peak sunspot numbers and the centers of sunspot latitudes ($C$). Furthermore, for stronger cycles, when sunspot eruptions start from higher latitudes, the overall latitude extents of sunspots are increased and thus the widths of distributions are increased. This explains the positive correlation between the peak sunspot numbers and $\delta$.

We remember that so far we have obtained all the parameters ($C$, $\delta$ and $H$) averaged over each cycle. These parameters, however, have variations within the cycle.
\citet{CS16} have studied the evolution of $C$ of the fitted Gaussians as functions of the cycle phase. 
They have shown that all cycles begin at different latitudes. Depending on the strengths, their activity levels first rise to highest values 
and then they all decay at the same rate; see their Fig.\ 3. 
To check whether our model reproduces this feature or not, we repeat the same analysis. 
We note that \citet{CS16} fitted the latitudinal distribution of sunspots with the Gaussian profile and obtained the center $C$. 
However, as discussed in \Sec{sec:evol}, the Gaussian does not statistically fit the latitude distribution, particularly for our model the fitting will be even worse. Therefore, 
we obtain the centers $C_{\rm{ moment}}$ directly from the latitude  distribution. 
We compute this from each one-year data for every cycle in each hemisphere.  
The variation of this $C_{\rm{ moment}}$ with the total number of spot in each year is shown 
in \Fig{model_cameron}. Thus in this plot, each curve represents the evolution of the center of the latitude distribution with the number of spot in each year. 
Every point in the plot corresponds to one-year data. It is obvious that cycle begin from the right bottom corner. Then with the progress of the cycle, spot number increases first and then decreases. The cycle eventually ends at the left bottom corner. The interesting features as emphasized in \citet{CS16} that different cycles begin at different latitudes, and depending on the cycle strength, they rise at different rates. In \Fig{model_cameron}, we see that stronger cycles (red and brown lines) begin at higher latitudes and they rise fast \citep[the Waldmeier effect---][]{Wald,KC11}, and they begin to fall earlier (at higher latitudes). Whereas weaker cycles (cyan and blue lines) rise slowly and they begin to fall later (at lower latitudes); compare peaks of all curves. Therefore, the most interesting fact is that, though different cycles rise differently (given by the Waldmeier rule), they all fall at the same rate.  
In summary, the evolution of the latitudinal distribution of our model spot-latitudes is in good agreement with the observed behavior as shown in \cite{CS16}.

\section{Conclusion and Discussion}\label{sec:summary}

We have analyzed the latitude distributions of sunspots 
over the past 16 solar cycles, which includes the newly digitized sunspot data from Schwabe and Sp{\"o}rer and RGO records. 
In our study, we define the cycle length carefully by excluding the hemispheric overlap that might occur during the solar minimum as described in \citet{CS16}. Thus in our study we capture the true hemispheric properties of the sunspot distributions which are free from any influence of the other hemisphere.
We show that the latitude distribution of sunspots do not statistically follow a Gaussian distribution. Distribution parameters, namely
$H$, $C$ and $\delta$ show significant positive correlations with the cycle strengths. 
These correlations are in general agreement with previous findings of \citet{2008A&A...483..623S},
although there are some differences in the values of the correlations as the cycle lengths are computed
differently in their study.
In general, the observed correlations imply that stronger cycles begin sunspot eruptions 
at relatively higher latitudes and they have wider extends of sunspot eruption latitudes.
Another interesting result from this study is the hemispheric asymmetry in the distribution parameters, specially in the center values ($C$). We find that the $C_N-C_S$ parameter oscillates around the zero value whereas the amplitude of this oscillation seems to decrease with the increase of cycle number.

To explore these features, we have applied a flux transport dynamo model
in which polar field at the solar minimum is varied randomly.
Although this model produces many features of the solar cycle correctly, the latitudinal distribution
of sunspots is not in a good agreement with observation because of fewer sunspots near the equator.
Despite of this dissimilarity, all correlations amongst the distribution parameters as obtained in observation are reproduced correctly. 
Our results are in broad agreement with that of \citet{2008A&A...483..623S} who reproduced similar correlations 
using a simplified thin shell dynamo model. While our model has a fixed single cell (in each hemisphere) meridional flow, \citet{2008A&A...483..623S} have shown that
the model without meridional flow and with meridional flow
produce slightly different results.
Our model also reproduces the detailed variation of these parameters within the cycle as demonstrated by \citet{CS16}. Thus our study supports a considerable variation in the polar field as the cause of the irregular solar cycle and variations in the latitude distributions.

The variation in the polar field is directly and indirectly observed in the Sun \citep{Muno13,Priy14}. 
This variation is primarily caused by the the scatter in the tilt angles of active regions around Joy's law \citep{Ca13,JCS14,HCM17,KM17}, 
which is actually observed in the Sun \citep[e.g.,][]{Das10,SK12}. 
While it is known that the variation in the meridional flow is expected to cause some variations in the solar cycle \citep[e.g.,][]{KC11,BD13}, 
due to difficulties in the measurements, we are uncertain about the amount of variation present in the deep meridional flow. However, the surface meridional flow is observed to change with the solar cycle and most of this variation is probably caused by the inflows around the active regions \citep{Gizon10}. This meridional flow perturbation can change the polar field from cycle to cycle \citep{Jiang10,HU14,STD15}. 
In our study, we have not considered how the polar field could be varied in cycle to cycle. This could be due to the observed 
tilt scatter around Joy's law and/or due to the variation in the surface meridional flow.

\begin{acknowledgements}
{We thank the reviewer for his/her invaluable suggestions which helped us to improve the quality of the paper. The authors also want to thank Mausumi Dikpati for providing valuable comments and suggestions. 
 SM and DB also thank the Science
\& Engineering Research Board (SERB) for the project grant
(EMR/2014/000626).
BBK is supported by the NASA Living With a Star Jack Eddy Postdoctoral Fellowship Program, administered by the University Corporation for Atmospheric Research. The National Center for Atmospheric Research is sponsored by the National Science Foundation.}
\end{acknowledgements}

 \bibliographystyle{apj}
 \bibliography{references}

\end{document}